\documentclass[showpacs,preprint,aps]{revtex4}
\usepackage{graphicx}
\usepackage{dcolumn}
\usepackage{bm}
\usepackage{amssymb} 
\usepackage{amsmath}
\begin{document}
\setcounter{page}{1}
\title 
{Comment on ``Quasinormal modes in Schwarzschild-de Sitter 
spacetime: A simple derivation of the level spacing of the frequencies"
}
\author
{D. Batic$^1$, N. G. Kelkar$^2$ and M. Nowakowski$^2$}
\affiliation{ 
$^1$ Dept. of Mathematics, Univ. of West Indies, Kingston 6, Jamaica\\
$^2$ Dept. de Fisica, Universidad de los Andes,
Cra.1E No.18A-10, Bogota, Colombia}
\begin{abstract}
It is shown here that the extraction of quasinormal modes (QNMs)  
within the first Born approximation of the scattering amplitude is 
mathematically not well founded. Indeed, the constraints on the existence of 
the scattering amplitude integral lead to inequalities for the imaginary parts  
of the QNM frequencies. For instance,
in the Schwarzschild case,  $0 \leq \omega_I < 
\kappa$ (where $\kappa$ is the surface gravity at the horizon) 
invalidates the poles deduced from the first Born approximation 
method, namely, $\omega_n = i n \kappa$. 
\end{abstract}
\maketitle 

In \cite{paper}, based on a scattering amplitude written in the first 
Born approximation, 
a simple derivation for the imaginary parts of the 
quasinormal mode (QNM) frequencies for Schwarzschild and Schwarzschild-de Sitter 
black holes is given. Similar results can also be found in \cite{visser}. 
For the sake of conciseness, we begin directly from Eqs (13) 
and (22) in \cite{paper} 
which give the scattering amplitudes in the first
Born approximation. 
In \cite{paper}, the authors claim that starting with the 
scattering amplitude $S(\omega)$ given in the first Born approximation, 
the quasi-normal modes can be found by computing the poles of 
$S(\omega)$. 

\section{Schwarzschild case} 
In the Born approximation and for the Schwarzschild case, the scattering 
amplitude is given as  
\begin{equation}
S(\omega)=\int_{2M}^{\infty}dr~U(r)e^{2i\omega
r_{*}(r)}=\int_{2M}^{\infty}dr~\left[\frac{\ell(\ell+1)}{r^2}+
\frac{2M}{r^3}\right]\biggl({r\over 2M} - 1\biggr)^{4iM\omega}e^{2i\omega
r}
\end{equation} 
with 
\begin{equation}
r_{*}=r+ 2M \ln{\biggl({r\over 2M} - 1\biggr)},\quad r>2M 
\end{equation} 
and 
\begin{equation}
\omega=\omega_R+i\omega_I,\quad \omega_R>0 \, , 
\end{equation} 
where $\omega_R$ and $\omega_I$ denote the real and imaginary parts 
of a QNM frequency, respectively. In particular, for $s=2,3$
we have to compute integrals of the form
\begin{equation}
I_s=\int_{2M}^{\infty}dr~r^{-s}\biggl({r\over 2M} - 1\biggr)^{4iM\omega}
e^{2i\omega
r}\propto\int_{0}^{\infty}dx~h(x)=\mathcal{I}_s,\quad
h(x)=\frac{x^{4iM\omega}}{(x+1)^s}~e^{4iM\omega x} \, , 
\end{equation} 
where we made the substitution $x=(r/2M) - 1 $. Taking into account
that
\begin{equation} 
|h(x)|\approx\left\{ \begin{array}{ll}
         x^{-4M\omega_I} & \mbox{as $x\to 0^{+}$},\\
         x^{-s-4M\omega_I}e^{-4M\omega_I x} & \mbox{as $x\to\infty$},\end{array}
         \right.,\quad s=2,3
\end{equation} 
the integral $\mathcal{I}_s$ will exist if the imaginary part of
the quasi-normal mode frequency is restricted to the following range
\begin{equation}\label{conditio}
0\leq\omega_I<\frac{1}{4M}.
\end{equation}
For $\omega_I\in[0,1/(4M))$ the above integral can be computed
exactly in terms of Whittaker's functions and Gamma functions as
follows \cite{GRAD}
\begin{eqnarray}\label{amplitude_exact} 
\mathcal{I}_s=C_1\Gamma\left(1+i\frac{\omega}{\kappa}\right)\Gamma\left(s-1-i\frac{\omega}{\kappa}\right)
M_{\frac{s}{2}+\frac{i\omega}{2\kappa},\frac{1-s}{2}+\frac{i\omega}{2\kappa}}\left(i\frac{\omega}{\kappa}\right)+
\\ \nonumber
C_2 \frac{\Gamma\left(1-s+i\frac{\omega}{\kappa}\right) 
e^{-{i\omega \over 2\kappa}}}
{1+s-i\frac{\omega}{\kappa}}\left[\frac{i s}{
-s+i\frac{\omega}{\kappa}}
M_{\frac{s}{2}+\frac{i\omega}{2\kappa},\frac{1+s}{2}-\frac{i\omega}{2\kappa}}\left(i\frac{\omega}{\kappa}\right)+
C_3
M_{1+\frac{s}{2}+\frac{i\omega}{2\kappa},\frac{1+s}{2}-\frac{i\omega}{2\kappa}}\left(i\frac{\omega}{\kappa}\right)\right] \, , 
\end{eqnarray}
where
\[
C_1= -{i \kappa \over \omega \Gamma(s)} \biggl( {i \omega \over \kappa}
\biggr)^{{s\over2} - {i \omega \over 2\kappa} } 
\quad
C_2= {\kappa \over \omega} \biggl( {-i \omega \over \kappa} \biggr) ^{{s\over 2} - {i \omega \over 2\kappa}},
\quad
C_3=\frac{\kappa(s+1)}{\omega }.
\]
Here, $\kappa=(4M)^{-1}$ is the surface gravity at the event
horizon and the Whittaker's function $M_{\mu,\nu}(z)$ is defined
as \cite{ABR}
\begin{equation}\label{whittaker}
M_{\mu,\nu}(z)=z^{\nu+1/2}e^{-z/2}\sum_{n=0}^{\infty}\frac{(\nu-\mu+1/2)_n}{n!(2\nu+1)_n}~z^n,\quad
(a)_n=\prod_{s=0}^{n-1}(a+s),\quad a\in\mathbb{C}.
\end{equation}
Note that the Whittaker function $M_{\mu,\nu}(z)$ is analytic
everywhere except at the points \cite{SLA}
\begin{equation}\label{condition}
2\nu=-1,-2,-3,\cdots \, , 
\end{equation}
where it has simple poles. Considering the Euler-Weierstrass
representation of the Gamma function
\begin{equation} 
\Gamma(z)=\frac{e^{-\gamma
z}}{z}\prod\left(1+\frac{z}{n}\right)^{-1}~e^{z/n}
\end{equation} 
(with $\gamma$ the Euler-Mascheroni constant) together with
(\ref{condition}) the poles of the amplitude $S(\omega)$ are
represented by the poles of the Gamma functions and of the
Whittaker's functions entering in (\ref{amplitude_exact}). These
poles have the general form, 
\begin{equation}\label{QNMSCH}
\omega_n=-in\kappa,\quad\omega_n=in\kappa,
\end{equation}
where the presence of a factor $n$ instead of $n+1/2$ is due to
the Born approximation which holds for $n\gg 1$, i.e. for high
energies. Therefore, all these poles have to be disregarded by 
virtue of the integrability condition (\ref{conditio}). Hence, the
conclusions drawn in \cite{paper,PAD,visser} concerning the large imaginary
parts of the QNM frequencies should be taken with some caution.
Note also that the QNMs come not only from the Gamma functions but also 
from the Whittaker functions. 

An alternative quick way to arrive at the same conclusion is to
observe that the integral $\mathcal{I}_s$ will pick up a
contribution only near the event horizon. In this case the problem
reduces to the computation of the integral
\begin{equation}\label{integrale}
\int_0^{\infty}dx~x^{i\frac{\omega}{\kappa}}e^{4iM\omega x}.
\end{equation}
The above integral belongs to the class of integrals FI II779 in
\cite{GRAD}
\begin{equation}\label{eq13}
\int_0^{\infty}dx~x^{\nu-1}e^{-\mu x}=\frac{\Gamma(\nu)}{\mu^\nu}
\end{equation} 
under the integrability condition
\begin{equation}\label{cond1}
{\rm{Re}}(\mu)>0,\quad {\rm{Re}}(\nu)>0.
\end{equation}
It is not difficult to see that for the integral (\ref{integrale})
the condition (\ref{cond1}) requires that
\begin{equation}
0<\omega_I<\kappa.
\end{equation} 
Since the integral (\ref{integrale}) exists if and only if
$\omega_{I}\in(0,\kappa)$ the Gamma function
$\Gamma(1+i\omega/\kappa)$ will not exhibit any pole for the
aforementioned range and the poles $\omega_n=in\kappa$ will lay
outside the range of validity of the method employed for the
search of the poles of the scattering amplitude. This means that
in the Born approximation the imaginary part of the quasi-normal
modes has to be restricted to the interval (\ref{condition}).

Before proceeding further we note that in physics one often introduces a 
regularization method when an integral does not exist. The Fourier transform
of the Coulomb potential (equivalent to the Born approximation of Coulomb
scattering) is a typical example where one introduces a regulator of the 
form $e^{-\mu r}$ in order to remedy the problem generated by the $1/r$ 
potential. In the present case, the integral (Eq. (\ref{eq13})) exists 
over a certain range of the parameter $\nu$ and hence in principle no 
regularization is required. Now if one decides nevertheless to perform 
a regularization for the sake of finding poles (though there is no 
physical justification for that), then one could write:
\begin{equation}
F = \lim_{a\to 0^+} \int_0^{\infty} t^{\nu -1} e^{-a/t} e^{-t} dt
= \lim_{a\to 0^+} 2 a^{\nu/2} K_{-\nu}(2\sqrt{a})), 
\end{equation}
where using 
$$
K_{\nu}(z) \propto {1 \over 2} \biggl (\Gamma(\nu) \biggl({z \over 2}
\biggr )^{-\nu}  
 \biggl (1 + {z^2 \over 4(1 - \nu)} + \, ...\biggr ) + 
\Gamma(-\nu) \biggl({z \over 2}\biggr )^{\nu} \, \biggl (1 + {z^2 \over 4(1 + \nu)} 
+ \, ...\biggr ) \biggr )
$$
with $z \to 0 \wedge \nu \not\in$ {\bf Z} (positive and negative integers), 
$F = \Gamma(\nu)$. 
However, $\nu$ cannot be a positive or negative
integer and therefore this invalidates again the extraction of the 
poles from the Gamma function above. 
Thus, even if one tries a regularization one cannot determine the poles.  
\section{Schwarzschild-de Sitter case}
According to \cite{paper} the tortoise coordinate and the scattering amplitude
for the Schwarzschild-de Sitter case will be given by 
\begin{equation}
r_{*}=\frac{1}{2\kappa_{-}}\ln{\left|\frac{r}{r_{-}}-1\right|}-\frac{1}{2\kappa_{+}}\ln{\left|1-\frac{r}{r_{+}}\right|}
-\frac{1}{2}\left(\frac{1}{\kappa_{-}}-\frac{1}{\kappa_{+}}\right)\ln{\left|\frac{r}{r_{-}+r_{+}}+1\right|} \, , 
\end{equation} 
where $r_{-}$ and $r_{+}$ are the event and cosmological horizons,  
respectively, and $\kappa_{+}$ and $\kappa_{-}$ the corresponding surface 
gravity terms. The scattering amplitude is now given as  
\begin{equation}
S(\omega)=\int_{r_{-}}^{r_{+}}dr~U(r)
\left(\frac{r}{r_{-}}-1\right)^{i\frac{\omega}{\kappa_{-}}}\left(1-\frac{r}{r_{+}}\right)^{-i\frac{\omega}{\kappa_{+}}}
\left(\frac{r}{r_{-}+r_{+}}+1\right)^{i\omega\left(\frac{1}{\kappa_{+}}-\frac{1}{\kappa_{-}}\right)}
\end{equation} 
with
\begin{equation} 
U(r)=\frac{\ell(\ell+1)}{r^2}+\frac{2M}{r^3}-\frac{2}{3r^2_\Lambda} \, ,
\quad r_{\Lambda} = {1 \over \sqrt{\Lambda}}. 
\end{equation} 
By means of the transformation $u=(r-r_{-})/(r-r_{+})$ mapping the
event horizon to $0$ and the cosmological horizon to $1$ we can
reduce the computation of $S(\omega)$ to the computation of the
following integral
\begin{equation} 
\mathcal{I}_s=\int_0^1 du F_s(u),\quad s=0,2,3 \, , 
\end{equation} 
where
\begin{equation}
F_s(u)=\left\{
\begin{array}{ll}
          \frac{u^{i\frac{\omega}{\kappa_{-}}}(1-u)^{-i\frac{\omega}{\kappa_{+}}}}{(1-yu)^s(1-zu)^{i\omega\left(\frac{1}{\kappa_{+}}-\frac{1}{\kappa_{-}}\right)}} & \mbox{for $s=2,3$},\\
         \frac{u^{i\frac{\omega}{\kappa_{-}}}(1-u)^{-i\frac{\omega}{\kappa_{+}}}}{(1-zu)^{i\omega\left(\frac{1}{\kappa_{+}}-\frac{1}{\kappa_{-}}\right)}} & \mbox{for $s=0$} \end{array}
         \right.
\end{equation} 
with
\begin{equation}\label{y,z}
y=\frac{r_{-}-r_{+}}{r_{-}},\quad
z=\frac{r_{-}-r_{+}}{r_{+}+2r_{-}}.
\end{equation}
Taking into account that
\begin{equation} 
|F_{s}(u)|\approx\left\{
\begin{array}{ll}
         u^{-\frac{\omega_I}{\kappa_{-}}} & \mbox{as $u\to 0^{+}$}~\Longrightarrow~\omega_I<\kappa_{-},\\
         (1-u)^{\frac{\omega_I}{\kappa_{+}}} & \mbox{as $u\to 1^{-}$}~\Longrightarrow~\omega_I>-\kappa_{+},\end{array}
         \right.,\quad s=0,2,3
\end{equation} 
it follows that the above integral exists if and only if the
imaginary part of the QNM frequency satisfies the condition
\begin{equation}\label{cond-Sds}
-\kappa_{+}<\omega_I<\kappa_{-}.
\end{equation}
Therefore, the poles of the scattering amplitude as computed in
\cite{paper}
\begin{equation}
\omega_n=in\kappa_{-},\quad\omega_n=-in\kappa_{+}\quad n\gg 1
\end{equation}
should be disregarded by  virtue of the integrability condition
(\ref{cond-Sds}).

In conclusion, we have shown that the limitation imposed by the requirement 
of the existence of the scattering amplitude integral impedes the 
deduction of the QNM frequencies using the Born approximation.


\begin{thebibliography}{999}

\bibitem{paper}
T Roy Choudhury and T Padmanabhan, Phys. Rev. D {\bf{69}}, 064033 (2004). 
\bibitem{visser}
A J M Medved, D Martin and M Visser, Class. Quant. Grav. {\bf{21}}, 1393 
(2004). 
\bibitem{GRAD}
I S Gradshteyn and I M Ryhzik, \textit{Table of Integrals,
Series and Products}, New York: Academic Press (1994). 
\bibitem{ABR}
M Abramowitz and I A Stegun, New
York: Dover Publications (1972). 
\bibitem{SLA}
L J Slater, \textit{Confluent Hypergeometric Functions},
Cambridge University Press (1960). 
\bibitem{PAD}
T Padmanabhan, Class.Quant.Grav.
{\bf{21}}, L1 (2003). 
\end{thebibliography}
\end{document}